\documentclass[a4paper, amsfonts, amssymb, amsmath, reprint, showkeys, nofootinbib, twoside, superscriptaddress]{revtex4-2}

\usepackage{hyperref}
\usepackage{graphicx}
\usepackage{amsthm, amsmath, amssymb}
\usepackage{tikz}
\usepackage{bbold}
\usepackage{footnote}
\usepackage{verbatim}
\usepackage{cancel}

\newcommand{\I}{\mathcal{I}}
\newcommand{\Stoi}{\mathcal{S}}

\begin{document}

\title{Robustness and complexity of directed and weighted metabolic hypergraphs}

\author{Pietro Traversa}
\affiliation{Institute for Biocomputation and Physics of Complex Systems (BIFI), University of Zaragoza, 50018 Zaragoza, Spain}
\affiliation{Department of Theoretical Physics, University of Zaragoza, 50018 Zaragoza, Spain}
\affiliation{CENTAI Institute, Turin, Italy}

\author{Guilherme Ferraz de Arruda}
\affiliation{CENTAI Institute, Turin, Italy}

\author{Alexei Vazquez}
\affiliation{Nodes \& Links Ltd, Salisbury House, Station Road, Cambridge, CB1 2LA, UK}

\author{Yamir Moreno}
\affiliation{Institute for Biocomputation and Physics of Complex Systems (BIFI), University of Zaragoza, 50018 Zaragoza, Spain}
\affiliation{Department of Theoretical Physics, University of Zaragoza, 50018 Zaragoza, Spain}
\affiliation{CENTAI Institute, Turin, Italy}

\begin{abstract}
Metabolic networks are probably among the most challenging and important biological networks. Their study provides insight into how biological pathways work and how robust a specific organism is against an environment or therapy.
Here we propose a directed hypergraph with edge-dependent vertex weight as a novel framework to represent metabolic networks. This hypergraph-based representation captures higher-order interactions among metabolites and reactions, as well as the directionalities of reactions and stoichiometric weights, preserving all essential information. Within this framework, we propose the communicability and the search information as metrics to quantify the robustness and complexity of directed hypergraphs.  We explore the implications of network directionality on these measures and illustrate a practical example by applying them to the small-scale e\_coli\_core model. Additionally, we compare the robustness and the complexity of 30 different models of metabolism, connecting structural and biological properties. Our findings show that antibiotic  resistance is associated with high structural robustness, while the complexity can distinguish between eukaryotic and prokaryotic organisms.
\end{abstract}

\maketitle

\section{Introduction}

A metabolic network~\cite{campbell_biology:_2018,dubitzky_metabolic_2013,small_world_network,stitt_metabolic_2010,metaboli_applications} is a highly organized system of chemical reactions that occur in living organisms to sustain life and regulate cellular processes. Metabolic networks are incredibly complex because of the large number of reactions and the intricate web of interactions between molecules. Chemical reactions take some metabolites, usually called reactants or substrates, and turn them into products, which can be used by other reactions.  
This complexity allows organisms to perform various functions and respond to various challenges, but it makes understanding them much more challenging. The key functions of metabolism are the production of energy, the conversion of food into building blocks of proteins, lipids, nucleic acids, and carbohydrates, and the elimination of metabolic wastes.

Given the network structure of metabolism, many researchers have attempted to characterize and understand it through network theory. It has been shown that graphs whose nodes are metabolites and are connected by chemical reactions have a scale-free distribution~\cite{small_world_network} and have been described as “among the most challenging biological networks and, arguably, the ones with most potential for immediate applicability"~\cite{guimera_functional_2005}. Other attempts have tried to give more concrete answers by focusing on graphs with reactions as nodes or bipartite graphs but missing a fundamental aspect of chemical reactions. To take place, they require a collective interaction of reactants to create multiple products. Hence, these are high-order interactions that the graphs cannot fully capture. As network theory has advanced, new structures have been devised that can capture high-order interactions. These structures called hypergraphs have been very successful in fields such as social sciences~\cite{Niu2017, Centola2018, Baronchelli2018, Benson-2018-simplicial, Iacopo2019, de_arruda_social_2020, landry2020effect, barrat2021social, FerrazdeArruda2021, Alvarez-Rodriguez2021, Leonie2020}, epidemiology~\cite{Bodo2016, de_arruda_social_2020, FerrazdeArruda2021,ferraz_de_arruda_multistability_2023, higham2021, higham2021b}, biology~\cite{Stewart2003, Golubitsky2005, Golubitsky06nonlineardynamics, Yu2011, Bairey2016, Petri2021, Cervantes-Loreto2021}, etc. Recently, Mulas et al.~\cite{mulas_spectral_2021,jost_hypergraph_2019} applied hypergraphs to chemical networks trying to capture the high-order nature of chemical reactions. In this paper, we take the concept of chemical hypergraphs and apply it to metabolic networks. In addition, we take it a step further by showing how including weights in the treatment allows no biological or structural information to be lost. Therefore, we argue that metabolic hypergraphs are the right framework to address and understand metabolism, allowing a bridge between biology and network theory. 

This article aims to lay the foundation for a theory of metabolic networks based on hypergraphs. We describe the method by which each metabolic network can be represented as a hypergraph and introduce two applicable measures, namely, communicability and search information. 

The work is organized as follows. In section~\ref{definitions} we give the mathematical definitions regarding metabolic hypergraphs. We also comment on previous studies in the field of metabolic networks and on how they can be viewed as a simplification of the metabolic hypergraph we propose here. In section~\ref{measures} we propose a generalization of the communicability and search information to hypergraphs. We keep this section general enough so that these measures can be easily applied to any hypergraph, directed or undirected, weighted or not. We use metabolic hypergraphs as an example and we report the results in section~\ref{results}. We conclude by commenting on the possibility that this framework offers to motivate further research in this area.

\section{Metabolic networks as hypergraph}\label{definitions}
In this section, we give a formal definition of metabolic hypergraphs and introduce the notation that is used to characterize them. 

\subsection{Hypergraphs definition}

A hypergraph $H=\left\{V, E \right\}$ is a set of vertices or nodes $v\in V$ and hyperedges $e \in E$. Each hyperedge is a subset of $V$ such that different nodes interact with each other if and only if they belong to the same hyperedge. Thus, unlike traditional graphs, where edges connect pairs of nodes, hyperedges represent interactions involving multiple nodes. If the dimension $|e|$ of the hyperedges is $2$, then the hypergraph is equivalent to a conventional graph. The total number of vertices is denoted as $N=|V|$ and the number of hyperedges as $M=|E|$.

To interpret metabolic networks as hypergraphs, we first need to define a special type of hypergraph introduced by Chitra et al.~\cite{chitra_random_2019}. A hypergraph with edge-dependent vertex weights (EDVW) $H=\left\{V,E,W,\Gamma \right\}$ is a set of vertices or nodes $v\in V$, hyperedges $e \in E$,  edge weights $w(e)$ and edge-dependent vertex weight $\gamma_e(v)$. If $\gamma_e(v)=\gamma(v) \; \forall \, e \in E$, then the hypergraph is said to have edge-independent vertex weight. All the weights are assumed to be positive. These types of weights are a unique property of some higher-order systems and are crucial to encode in the hypergraph all the information contained in metabolic networks. 

In this paper, we deal with directed hypergraphs, which are an extension of directed graphs. In a directed hypergraph, each hyperedge is associated with a direction similar to the direction of an arrow connecting two vertices in a directed graph. In this context, a hyperedge $e_j$ is divided into a head set $H(e_j)$ and a tail set $T(e_j)$. Similarly to the arrow, the direction goes from the tail to the head set, with the difference that the directed hyperedge is connecting multiple vertices. A vertex can belong solely to either the head or the tail of a hyperedge, but not both. Unless explicitly stated otherwise, any hypergraph in this paper is considered to be directed.

Additionally, we define $k_v^{out}$, the out-degree of a vertex $v \in V$, as the number of hyperedge-tails that include $v$. Similarly, $k_v^{in}$ denotes the in-degree of a vertex $v \in V$, the number of hyperedge-heads in which $v$ is contained. We also use $|H(e)|$ and $|T(e)|$ to represent the number of vertices belonging to $H(e)$ and $T(e)$ respectively.

Given a directed hypergraph H = $\left\{ V,E\right\}$ of $N$ vertices and $M$ hyperedges, the incidence matrix is the matrix $\mathcal{I} \in \mathbb{R} ^{N\times M}$ such that:
    \begin{equation}\label{Directed Incidence}
        \mathcal{I}_{ij} = \begin{cases}
        1 &\text{ if $v_i \in H(e_j)$}\\
        -1 &\text{ if $v_i \in T(e_j)$}\\
        0 &\text{ if $v_i \not\in e_j$}
        \end{cases},
    \end{equation}
where $ H(e_j)$ and $ T(e_j)$ are, respectively, the head and the tail of the hyperedges $e_j$. We can rewrite the incidence matrix as 
\begin{equation}
    \I = \I_H - \I_T,
\end{equation}
where we separated the contributions coming from the head and the tail of the hyperedges in order to work with positive signed matrices.

\subsection{Metabolic hypergraphs}\label{Sec: metabolic hypergraph}

In this article, we focus on metabolic networks. A metabolic network \cite{dubitzky_metabolic_2013} is a set of biological processes that determines the properties of the cell. Several reactions are involved in metabolism, grouped into various metabolic pathways. A metabolic pathway is an ordered chain of reactions in which metabolites are converted into other metabolites or energy. For example, the glycolysis pathway is the set of reactions involved in the transformation of one molecule of glucose into two molecules of pyruvate, producing energy.
Metabolic networks are among the most challenging and highest potential biological networks \cite{guimera_functional_2005,small_world_network}. The way to represent a metabolic network on a graph is not unique, and several approaches have been tried. One possible way is to consider metabolites (or reactions) as nodes and connect them if and only if they share a reaction (or metabolite). The resulting graph is undirected, and this may change the structural properties of the network in an undesirable way. In \cite{beguerisse-diaz_flux-dependent_2018}, the authors analyze the same dataset that we analyze for E.Coli and propose a directed graph with reactions as nodes that take into account the directionality of the reactions, highlighting the difference with the undirected counterparts.

\begin{figure*}
    \centering
    \includegraphics[width=0.95\textwidth]{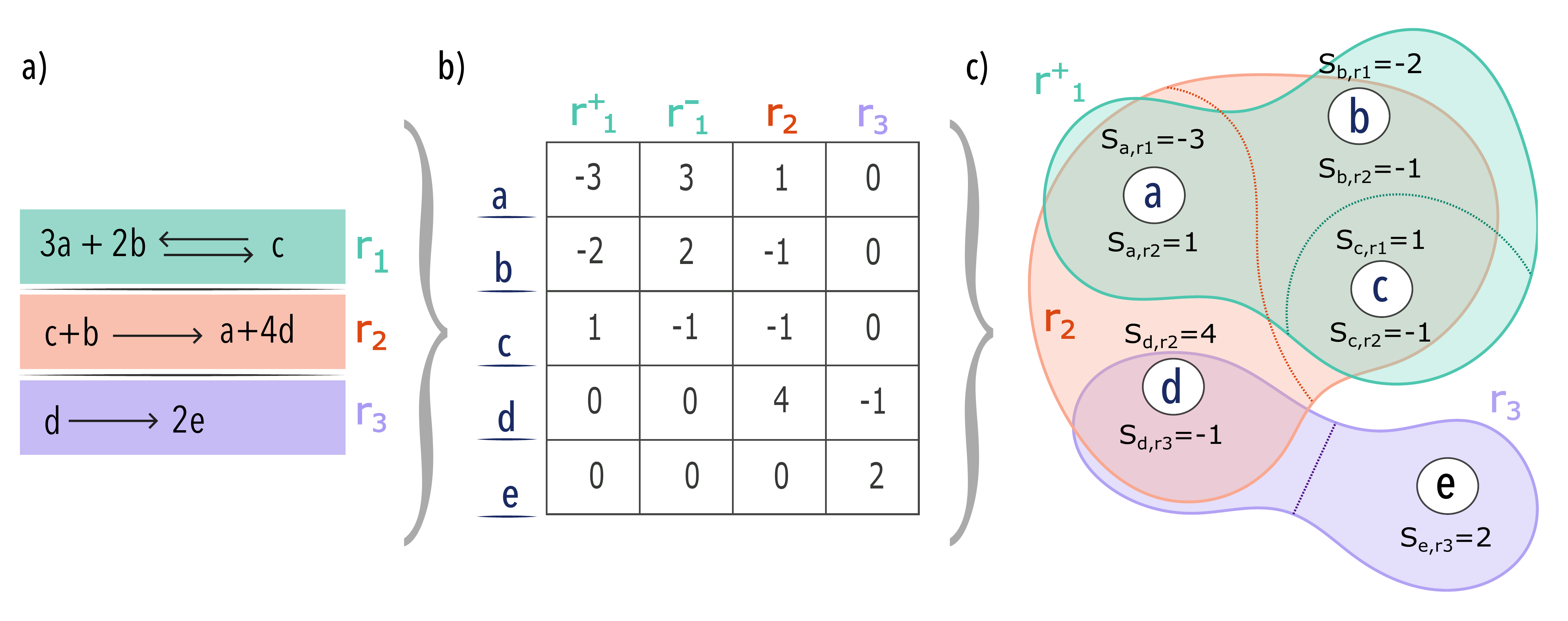}
    \caption{An example of a metabolic network mapped into a hypergraph with edge-dependent vertex weight. In a), we present a small network composed of three reactions and five metabolites. The first reaction $r_1$ is reversible and is represented with the double arrow. In b), we show the corresponding stoichiometry matrix. Reactants are negative and products are positive. Note that we need to split the reversible reaction into two irreversible reactions $r_1^+$ and $r_1^-$ to write it in matrix form. This stoichiometry matrix is the weighted incidence matrix of the hypergraph with edge-dependent vertex weights shown in c). For the sake of visualization, only the hyperedge $r_1^+$ is shown. The hyperedge $r_1^-$ is just the same but with the opposite sign. Note that weights are both positive and negative, meaning that the hypergraph is directed. Indeed, we separate the head and tail of each hyperedge with a dashed line.}
    \label{Fig: met}
\end{figure*}

However, reactions are intrinsically higher-order interactions since they can occur only when all reactants are present. In Fig. \ref{Fig: met}, we illustrate the way to map a chemical reaction network into a hypergraph. The resulting hypergraph is a directed hypergraph with edge-dependent vertex weight, which we will refer to as metabolic hypergraph for brevity. 
More formally, we define a metabolic hypergraph as a 3-tuple $H = \left\{V, E, \Stoi\right\}$, where $V = \{v_1,v_2,\dots v_N\}$ is a set of N metabolites (vertices), and E is a set of oriented reactions (hyperedges). Each $e\in E$ is a pair $(T(e), H(e))$, the tail and the head of the hyperedge which corresponds respectively to the inputs and outputs of the reaction. Note that $T(e)$ or $H(e)$ can also be empty sets. This is the case for external reactions that introduce inside the cell the ingested metabolites (the tail is an empty set), and external reactions that secrete metabolites (the head is an empty set). We also call the former source reactions and the latter sink reactions, and their effect on the measurements is discussed in more detail in section \ref{measures}. $\Stoi$ is the stoichiometry matrix associated with the chemical network and it represents the EDVW of the hypergraph. Indeed, one can notice that $S$ can be rewritten using the EDVW matrix $\Gamma$ as $\Stoi = \Gamma \circ \I$, where $\I$ is the directed incidence matrix and “$\circ$” is the element-wise matrix product.

\subsection{Literature background}

There are different techniques to study metabolic networks. One popular method is using stochastic chemical kinetics~\cite{gillespie1977exact_kynetic}, but this requires the notion of the kinetic rates constant, the rates at which metabolites are consumed per reaction, which are usually not available~\cite{Kynetic_limitation_srinivasan2015}. What instead is generally known are the reactions, the stoichiometry coefficient, and the structure of the metabolic network. Thus, several graph representations of metabolic networks have been tried. The most common one is the reaction adjacency matrix (RAG) defined as $A^{RAG} = \hat{\Stoi}^T \hat{\Stoi}$~\cite{beguerisse-diaz_flux-dependent_2018}, where $\hat{\Stoi}$ is the boolean version of the stoichiometry matrix. The biggest limitation of this model is that is undirected, while we know that the direction of reactions is chemically really important. A big improvement was proposed in~\cite{beguerisse-diaz_flux-dependent_2018} where the authors proposed a flux-dependent graph model, that accounts suitably for the directness of the reactions. However, graph representations of these systems are still missing a crucial point, which is the fact that reactions are higher-order object, that involves the interactions of all input metabolites to produce output metabolites. Therefore, hyperedges are the natural mathematical object to encode reactions. Mulas et al.~\cite{jost_hypergraph_2019} already took a step in this  direction by defining a Laplace operator for chemical hypergraphs. The last step we make is to incorporate into the hypergraph model the weights associated with metabolites and reactions, using a similar framework to the EDVW defined in~\cite{chitra_random_2019}. This last modification to the model is crucial to include biological and chemical constraints into the model.

The great advantage of the metabolic hypergraph framework we propose is that it captures all the physical properties that a metabolic network displays: the directness of reactions, the higher-order interactions, and the chemical properties like mass conservation, thanks to the inclusion of weights. This framework represents a link between network theory and biology.

We remark that the previous graph representation of metabolic networks can be seen as a pairwise projection of a metabolic hypergraph. For example, the RAG is an undirected projection of the hypergraph as in~\cite{traversa2023unbiased} and the flux-dependent graph~\cite{beguerisse-diaz_flux-dependent_2018} is similar to the normalized adjacency matrix defined in~\cite{BANERJEE_normalized} but extended to directed and weighted hypergraphs. Projections are a pairwise simplification and can perform well depending on the task, but they don't contain all the information.

\subsection{Dataset}

In our experiments, the metabolic hypergraphs are taken from the BiGG Database~\cite{BiGG_ref}. We analyze 30 different models, with an increasing number of nodes describing different organisms (see Table~\ref{tab:BiGG_table} in the Appendix for the exact number of nodes and reactions of each BiGG model). We chose the metabolic networks in order to have a reasonable variety of organisms, and we avoided very large networks because of the computational costs. The majority of the data is composed of bacteria that can be divided into classes like antibiotic-resistant, aerobic or anaerobic, Gram-positive or Gram-negative. The other organisms are Eukaryotes and one in the Archaea domain. All data are publicly available on the BiGG models web page~\cite{BiGG-models} in different formats. In this analysis, the \emph{.json} format is used. The data contain information on metabolites, reactions, and genes. Metabolites are identified by a Bigg id, consisting of an abbreviation defining their type, for example, “h” for hydrogen and “ATP” for the adenosine triphosphate, and a subscript indicating the compartment to which they belong. Regarding the reactions, in addition to their IDs, the metabolites belonging to them are given, with their respective stoichiometric coefficients. We work in the convention in which a metabolite with a positive stoichiometric coefficient is a product, otherwise, it is a reactant. In the BiGG dataset, the direction of the reactions is also determined by the parameters “lower\_bound” and “upper\_bound.” These parameters are associated with each reaction and correspond to the maximal flux of metabolites that can flow through. A value of lower\_bound $=0$ and upper\_bound $>0$ means that the reaction is annotated correctly, following the convention. On the contrary, if lower\_bound $<0$ and upper\_bound $=0$, the reactions are written with inverted orientations. These two parameters combined also determine if a reaction is reversible or not. If a reaction is reversible, both the direct and inverse reactions are present and will be characterized by a lower\_bound $<0$ and upper\_bound $>0$. We recall that we treat reversible reactions as two distinct hyperedges, see Fig.~\ref{Fig: met} for a visual example.
It is important to notice that few reactions have lower\_bound $=0$ and upper\_bound $=0$. In practice, this implies that no flux of metabolites can flow through, so those reactions are discarded. 

Lastly, we highlight that some hyperedges may have an empty tail or head. These hyperedges correspond to reactions involved in the transportation of metabolites from the outside of the cell to the inside or vice-versa. For this reason, sometimes they may represent sinks and sources in the hypergraph. By source, we mean a node or hyperedge from which you can start and leave but never go back, while a sink is a trapping node or hyperedge that if it is reached, it is impossible to leave.

\section{Measurements}\label{measures}

In this section, we define two measures of the chemical hypergraph based on the notion of paths or walks on hypergraphs. A \emph{walk} of length $l$ from node $v_0$ to node $v_l$ is defined as a sequence of alternating nodes and hyperedges $\left(  v_0,e_1,v_1,e_2,v_2, ... e_l,v_l \right)$. We also define the \emph{dual walk} form hyperedge $e_0$ to hyperedge $e_l$ of length $l$ as the alternating sequence of alternating nodes and hyperedges $\left(  e_0,v_1,e_1,v_2,e_2, ... v_l,e_l \right)$. We are interested in both metabolites and reactions, which is why it is useful also to consider the dual walk.

\subsection{Hypergraph communicability}

We are usually interested in understanding how paths are distributed because that is how information and interactions spread. In social systems, for example, the more the paths connecting two nodes, the easier is for information to spread from one another. Also, if one path of connection fails, the information can still be spread through other paths, even if they are longer than the path that failed. For this reason, the notion of paths and communication between nodes can also be related to the robustness of the network. However, having a robust network is not always positive. The same reasoning about the spreading of information applies to the spreading of viruses. If a network is robust, is way more difficult to design containment strategies for the virus, since shutting down a connection might not be enough because of the presence of alternative paths. A way to measure how nodes communicate within a network is called communicability and we extend this definition to hypergraphs.

The communicability~\cite{estrada_physics_2012,estrada_communicability_1} between two pairs of node $p$ and $q$ is defined as the weighted sum of all walks starting from node $p$ and ending at node $q$, as in
\begin{equation}
    G_{pq} = \sum_{k=0}^{\infty} c_k n_{pq}^k,
\end{equation}
where $n_{pq}^k$ is the number of walks from $p$ to $q$ and $c_k$ is the penalization for long paths. The most common choice is $ c_k=\frac{1}{k!}$ so that you recover an exponential expansion. For a graph, $n_{pq}^k$ can be easily found by taking the k-power of the adjacency matrix, $(A^k)_{pq}$. Hypergraphs don't have a unique definition of adjacency matrix, we thus have to use the definition of walk given above.
The vertex-to-vertex communicability for a hypergraph with incidence matrix $\I$ is defined as 
\begin{equation}
    G_{pq}^V = \sum_{k=0}^{\infty} \frac{\left((\I_T \I_H^t)^k\right)_{pq}}{k!},
\end{equation}
or in matrix form 
\begin{equation}
    G^V = e^{\I_T \I_H^t},
\end{equation}
where $t$ indicates the transpose of the matrix.
In metabolic hypergraphs, we are also interested in how reactions communicate with each other. For this reason, we define the hyperedge-to-hyperedge communicability based on the notion of dual path,
\begin{equation}
    G_{pq}^E = \sum_{k=0}^{\infty} \frac{\left((\I_H^t \I_T)^k\right)_{pq}}{k!},
\end{equation}
or in matrix form 
\begin{equation}
    G^E = e^{\I_H^t \I_T}.
\end{equation}

The Estrada index \cite{estrada_physics_2012,estrada_index_ref} of a hypergraph $H$ is generalized as
\begin{equation}\label{eq: EE}
\begin{split}
    EE^V(H) = \text{Trace}\left( G^V\right),\\
    EE^E(H) = \text{Trace}\left( G^E\right).
\end{split}
\end{equation}
One can notice that the matrices $\I_T \I_H^t$ and $\I_H^t \I_T$ have the same spectrum except for the number of zero eigenvalues because of the difference in size. This means that for $M>N$ for example (which is usually the case in metabolic hypergraphs), then the Estrada index defined on nodes and the one defined on the hyperedges are related by $EE^E(H)=EE^V(H)+(M-N)$.
We use the Estrada index defined on the nodes to measure the hypergraph robustness, also known as natural connectivity, as
\begin{equation}
    \Bar{\lambda}^V = \log{\left(\frac{EE(H)^V}{N}\right)}.
\end{equation}
The same definition holds for $\Bar{\lambda}^E$ with the proper normalization.

Since computing the exponential of very large matrices might be a difficult numerical task, we use an approximation for the calculation of the robustness based on eigenvalue decomposition. For simplicity, let us call $A^V = \I_T \I_H^t$ (the same reasoning holds for $A^E = \I_H^t \I_T^t$) and order the spectrum of $A^V$ in such a way that $\lambda_1>\lambda_2>\lambda_3>... \lambda_N$. Then the natural connectivity or robustness of the hypergraph becomes 

\begin{align*}
    \Bar{\lambda}^V & = \log\left(\sum_{i=1}^{N}e^{\lambda_i}\right) - \log(N) = \\
    & = \log\left[e^{\lambda_1}\left(1 + \sum_{i=2}^{N}e^{\lambda_i-\lambda_1}\right)\right] - \log(N) =\\
    & = \lambda_1 + \log\left(1 + \sum_{i=2}^{N}e^{\lambda_i-\lambda_1}\right) - \log(N) =\\
    & = \lambda_1 - \log(N) + \mathcal{O}\left(e^{-(\lambda_1 - \lambda_2)}\right)  .
\end{align*}
Thus if the spectral gap is large enough, the natural connectivity is dominated by the largest eigenvalue. Since the correction is exponential, this approximation is usually quite good. As a consequence of the common spectrum of $\I_H^t \I_T$ and $ \I_T \I_H^t$, the difference in robustness is approximately $\Bar{\lambda}^V - \Bar{\lambda}^E \approx \log(\frac{M}{N})$, which is usually quite small.

This generalization of communicability applies also to undirected hypergraphs by substituting $I_H$ and $I_T$ with the undirected incidence matrix $I$.

\subsection{Hypergraph search information}

\emph{Rosvall et al.}~\cite{rosvall_networks_2005,hide_and_seek} introduced the concept of search information, as a measure of complexity in urban graphs. The idea is to measure the number of binary questions one has to make in order to locate the shortest path connecting a node $s$ to a node $t$. As a consequence, this measure is based on walks like the communicability, but with the crucial difference that it considers only the shortest paths. This allows us to link the search information with the notion of complexity. While alternative pathways tend to make the network more robust, they also make the probability of finding the shortest path decrease and the complexity increases. This trade-off is the reason that motivated us to consider communicability and search information together.

In \cite{rosvall_networks_2005}, the search information is defined as a matrix $S$ with entries
\begin{equation}
    S(i,j)^V = -\log_2 \left( \sum_{\left\{p(i,j)\right\}} P\left(p(i,j)\right)\right),
\end{equation}
where $\left\{p(v_i,v_j)\right\}$ is the set of all shortest paths from node $v_i$ to node $v_j$.

The original definition was made for undirected and unweighted ordinary graphs, so a very different structure from directed hypergraphs with edge-dependent vertex weight but the meaning remains the same. What changes, is the probability of following the shortest path.
The probability of making a step is proportional to the stoichiometric coefficients of the starting and arriving node, similar to what has been done in the normalized flow graph in~\cite{beguerisse-diaz_flux-dependent_2018}. The probability of taking a step in a directed hypergraph with EDVW is
\begin{equation}
    \begin{split}
    P(v \xrightarrow{}e)= \frac{\gamma_e(v)}{\sum_h \gamma_h(v)},\\
    P(e \xrightarrow{}v)= \frac{\gamma_e(v)}{\sum_n \gamma_e(n)}.
\end{split}
\end{equation}
The probability of following a path is derived by multiplication of the single-step probability,
\begin{equation}
    P(v_0,v_l) = P(v_0 \xrightarrow{}e_1)P(e_1 \xrightarrow{}v_1) \dots P(e_l \xrightarrow{}v_l).
\end{equation}

It is important to note that the search information might be ill-defined if the hypergraph has sources or sinks. For example, by definition, there are no paths from a sink node $v_{\text{sink}}$ to any other nodes $v$, making the definition of $S(v_{\text{sink}},v)$ unclear in this case. What we do to solve the problem is to set $S(v_{\text{sink}},v)=0$ and then not count sink and source nodes when computing the average. With this convention, the access, hide, and average search information are defined as
\begin{equation}
\begin{split}
    & A^V(s) = \frac{1}{N-N_{\text{sources}}}\sum_t S^V(s,t)\\
    & H^V(t) = \frac{1}{N-N_{\text{sinks}}}\sum_s S^V(s,t)\\
    & \bar S^V = \frac{1}{(N-N_{\text{sinks})}(N-N_{\text{sources}})}\sum_{s,t} S^V(s,t) .\\
\end{split}
\end{equation}
As a consequence, the access information of a sink and the hide information of a source will be set to zero.
Following~\cite{rosvall_networks_2005}, we introduce an additional normalization factor $\ln_2 M$ to take into account size effects. We denote the normalized average search information as $\sigma^V=\frac{\bar S^V}{\log_2 N}$.
The interpretation of these measures is very intuitive. The access information measures how easy it is to reach the other nodes in the network, while the hide information estimates how hidden a node is. Consequently, very central and connected nodes in the hypergraph have low hide information because there are a lot of paths leading to them but have relatively high access information because there are also many paths departing from such nodes. 

\begin{figure*}[t!]
    \centering
    \includegraphics[width=\textwidth]{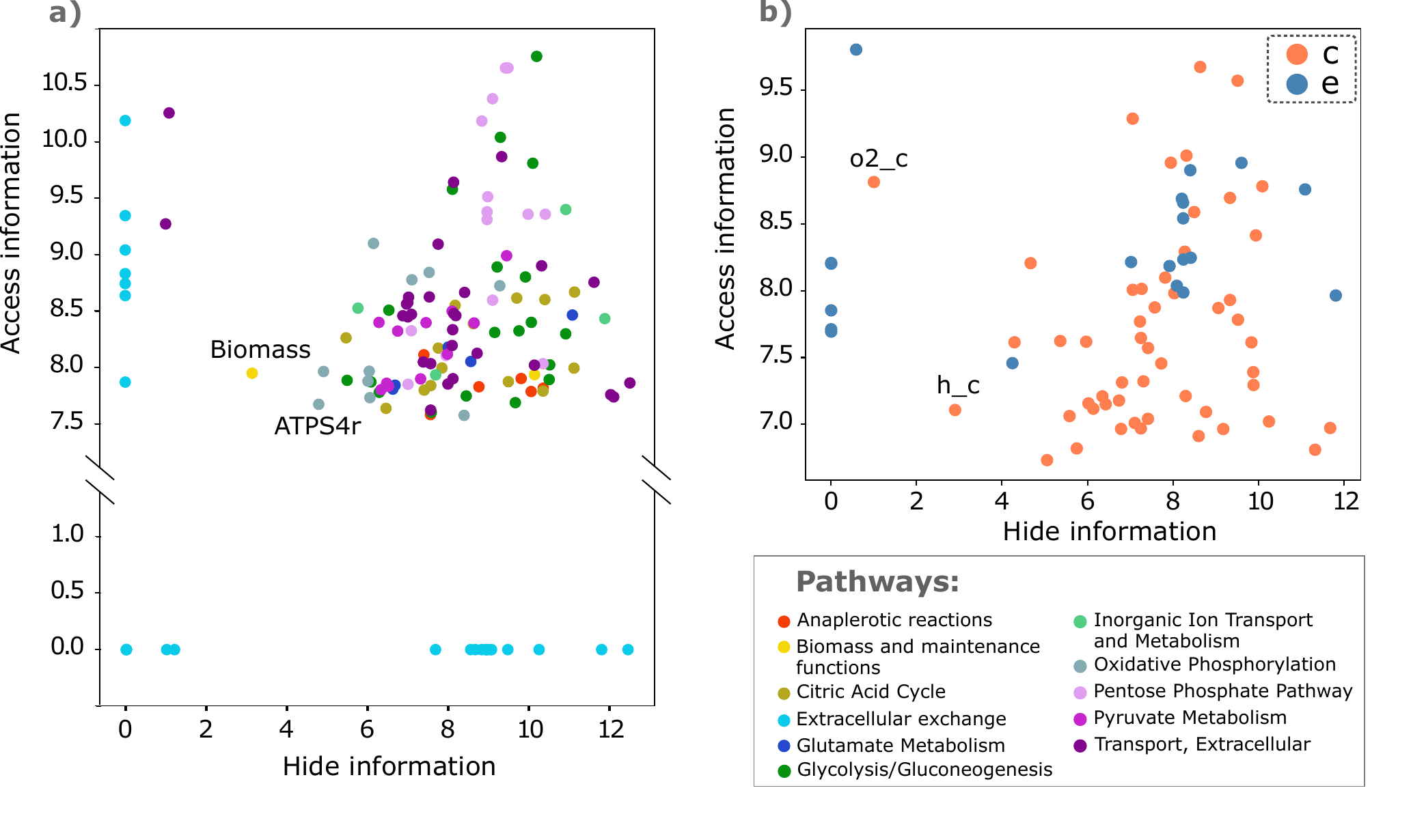}
    \caption{Access vs. hide information for reactions a) and metabolites b). Reactions are colored differently according to the pathway they belong to. Not that the $y$ axis is cut for visualization purposes. Metabolites are divided into compartments, $c$ stands for cytosol compartment, and $e$ for extracellular space.}
    \label{Fig: e_coli_core_example}
\end{figure*}

\section{Results and Discussion}\label{results}
In this section, we apply the previously defined metrics to a range of metabolic hypergraphs. As illustrated in Fig.~\ref{Fig: met}, these hypergraphs were constructed by starting with metabolic networks obtained from the BiGG Dataset~\cite{BiGG_ref}. The metabolic networks were selected to have a reasonable variety of organisms. 
The primary goal of this section is to demonstrate the practical application of our framework and the defined measurements.

\subsection{Exploring the E. coli Core Model: A Practical Example}

To provide a tangible illustration of our methodology, we focus on the BiGG model known as e\_coli\_core~\cite{BiGG_e_coli_core}. This model represents a small-scale version of Escherichia coli str. K-12 substr. MG1655,  making it an ideal candidate for demonstrating the performance of our metrics and understanding their limitations. Additionally, an Escher map for this model is available online~\cite{e_coli_fba_map}.

In Fig.~\ref{Fig: e_coli_core_example}, we show the access vs. hide information for reactions and metabolites.
Regarding the reactions (Fig.~\ref{Fig: e_coli_core_example} a), the measure correctly identifies the Biomass reaction as a central hub. Reactions are plotted with different colors based on the biological pathway they belong to. We can clearly see the behavior of sinks and sources in the reactions belonging to the extracellular exchange pathway. The pathways don't tend to separate into clusters, indicating that they all have a similar complexity. This could be an effect of the simplicity of this model or could be a property shared by all organisms. We didn't investigate further since the scope of this section was just to provide a practical example, but it could be worth it to explore it in future work. 

\begin{figure}[t!]
    \centering
    \includegraphics[width=\linewidth]{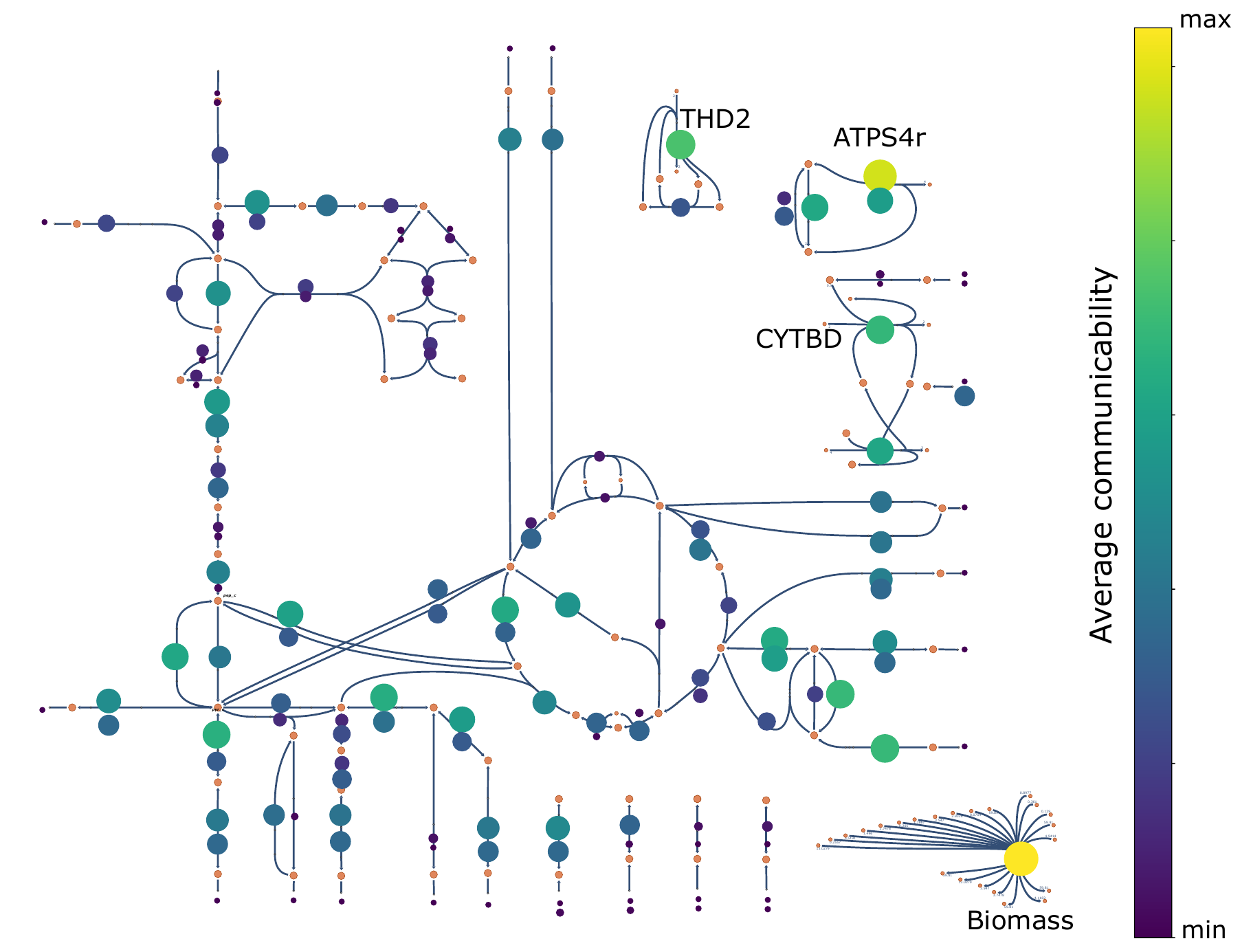}
    \caption{Reactions average communicability for the e\_coli\_core model. A simplified Escher map is used as a background to help with the visualization. For a more accurate version of the map, visit~\cite{e_coli_fba_map}.}
    \label{Fig: communicability_e_coli}
\end{figure}

We also comment on the reactions that are ranked the highest by the average communicability. The average communicability is defined as $\Bar{G}_e = \frac{1}{M}\sum_{h \in E} G^E_{he}$ and is shown in Fig.~\ref{Fig: communicability_e_coli}. Notably, the Biomass reaction (1-st highest average communicability) and ATP synthase (2-st highest average communicability) are correctly identified as central reactions within the metabolism. The Biomass reaction is responsible for cell growth, while ATP synthase plays a crucial role in ATP synthesis, the primary energy source for the organism. 
The production of ATP is mainly due to the consumption of oxygen that occurs through the reaction CYTBD (cytochrome oxidase bd -- 6-th highest average communicability). When oxygen is unavailable, Escherichia coli can still survive thanks to the activation of the anaerobic pathway, which derives energy from the reaction THD2 (NAD(P) transhydrogenase --3-rd highest average communicability). 

Regarding the metabolites (Fig.~\ref{Fig: e_coli_core_example} b), we observe a clear distinction between those belonging to the cytosol compartment and those located in the extracellular compartment. As expected, extracellular metabolites tend to have, on average, higher hide information. It is important to clarify that metabolites with zero hide information are source nodes and remain initialized to zero because they are unreachable. However, an instructive observation could be made on o2\_c. As commented in section~\ref{measures}, a node with low but non-zero hide information is expected to be a central hub, but in reality, it has a very low degree. The explanation for this helps to understand the implications of network directionality. The node o2\_c is only connected to the core metabolism via the irreversible CYTBD reaction as a substrate. Consequently, there cannot be any directed path from the core metabolism to o2\_c, only the opposite. We conclude that the node o2\_c does not belong to the largest strongly connected component. In practice, it behaves very similarly to a source node. Nonetheless, the hide information is not zero because a pathway originates from the transport of external oxygen to the cytosol. In contrast, in cyanobacteria, algae, and plants (not investigated here) O2 is produced via oxygenic photosynthesis. In those organisms, O2 should be part of the strongly connected component.

\subsection{Robustness and complexity across organisms}

Our study assesses the robustness and complexity of 30 distinct metabolic hypergraphs  derived from various eukaryotic and prokaryotic organisms.

\begin{figure*}[t!]
    \centering
    \includegraphics[width=0.95\textwidth]{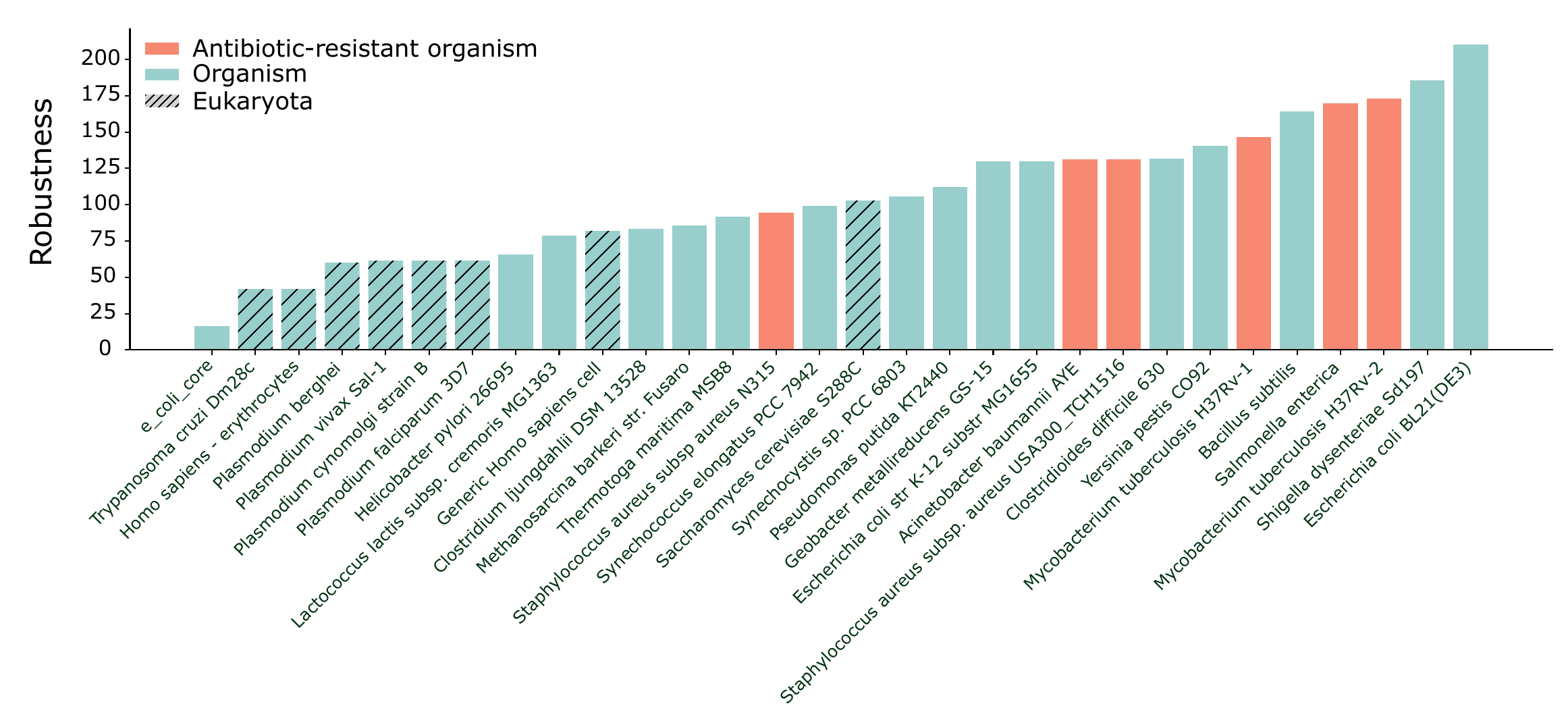}
    \caption{The robustness measured as the natural connectivity $\bar \lambda^V$ of 30 different BiGG models. The organisms resistant to antibiotics are shown in different colors. The models are ordered in increasing robustness.}
    \label{Fig: Robustness Compared}
\end{figure*}

In Fig.~\ref{Fig: Robustness Compared}, we present the computed robustness values for several organisms arranged in ascending order. The BiGG model associated with the organisms \emph{Staphylococcus aureus subsp aureus}~\cite{bigg_staphylococcus_N315_antibiotic,bigg_staphylococcus_USA300_antibiotic}, \emph{Mycobacterium tuberculosis}~\cite{bigg_tuberculosis_1_antibiotic,bigg_tuberculosis_2_antibiotic}, \emph{Acinetobacter baumannii AYE}~\cite{bigg_acinetobacter_antibiotic}, and \emph{Salmonella enterica}~\cite{bigg_salmonella_antibiotic} are represented in different color because they are bacteria that have evolved resistance to antibiotics. Except for the first \emph{Staphylococcus aureus subsp aureus} model, antibiotic-resistant bacteria tend to exhibit relatively high robustness compared to other organisms. We measured the Spearman's rank correlation between robustness and antibiotic resistance obtaining a value of $0.424$, revealing a moderate correlation. Here, the definition of robustness is based on the network's resilience to random or targeted node removal. The concept of natural connectivity quantifies this resilience by counting the number of closed loops in the network. If there are many alternative paths, it is less probable that a node removal will disconnect the network. 
In the context of biology, antibiotics operate by targeting and inhibiting some specific reactions, without which the cell dies~\cite{campbell_biology:_2018}. Therefore, having a structurally robust metabolism is advantageous as it allows the organism to circumvent antibiotic inhibition by utilizing alternative reactions or pathways. However, this is not the whole picture since many other factors play a role. For example, bacteria are naturally subjected to random mutations that may strengthen their response to antibiotics, and this may not necessarily be reflected in a high structural hypergraph robustness. Conversely, a very robust metabolic hypergraph, with many alternative paths, may have a few but very important reactions that are easy to target with antibiotics. Hence, high structural hypergraph robustness does not guarantee antibiotic resistance.  

\begin{figure*}[t!]
    \centering
    \includegraphics[width=0.95\textwidth]{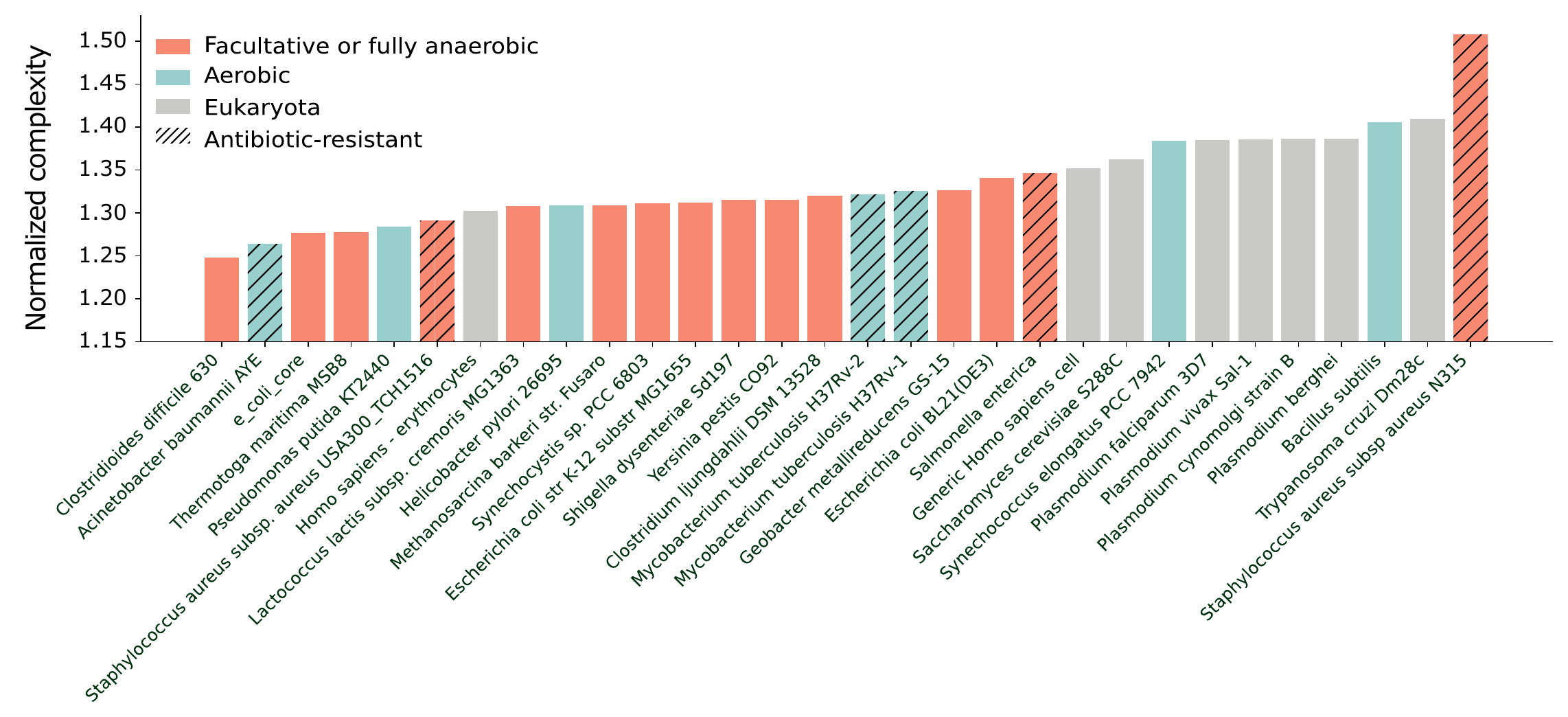}
    \caption{The complexity measured as the average search information $\sigma^V = \frac{S^V}{\log_2N}$ of 30 different BiGG models. The models are ordered in increasing complexity, and the y-axis is zoomed in for visualization purposes.}
    \label{Fig: Met complexity compared}
\end{figure*}

The complexity of metabolic networks is expected to be quite similar across organisms since they share many common reactions and metabolic pathways. Nevertheless, some differences are expected in the metabolism of aerobic and anaerobic organisms, as well as between eukaryotes and prokaryotes. Aerobic and anaerobic organisms should have a different metabolism because of the different ways they produce energy, while eukaryotes and prokaryotes have significantly different cell structures. 
With this in mind, we measure the average search information of the 30 different metabolic hypergraphs and report the results in Fig.~\ref{Fig: Met complexity compared}. We notice a clear separation between eukaryotes and some aerobic organisms, showing a high complexity, and prokaryotes, which have a lower complexity. A few outliers exist, including \emph{Staphylococcus aureus subsp aureus N315}, which exhibits high complexity, potentially due to unusually large weights associated with certain reactions compared to other organisms. Setting all the weights to $1$ would indeed lead to a much lower complexity, ranked slightly below the average, indicating a possible bias. In addition, one can also notice that the other model for \emph{Staphylococcus} has a low complexity. 
Another outlier is the first model we analyzed for \emph{Homo sapiens - erythrocytes}~\cite{bigg_homo_sapiens} that may be expected to be complex. However, it is important to note that this model refers just to the erythrocyte metabolism (blood cells) rather than the entire human metabolisms. Erythrocytes lack mitochondria and produce ATP through anaerobic glycolysis, so their metabolism could be closer to that of anaerobic organisms. Conversely, the low complexity of the aerobic organisms \emph{Acinetobacter baumannii AYE}, \emph{Pseudomonas putida}, and \emph{Helicobacter pylori} is curious, and we don't have a clear motivation. Note that a generic human (\textit{Homo sapiens}) cell has a similar complexity as a yeast cell (\textit{Saccharomyces cerevisiae}). That is expected. Eukaryote cells have similar metabolic pathways. The additional complexity in human metabolism is due to the multi-cellularity, which is not accounted for in this study.
\vspace{-0.5cm}
\section{Conclusion}

Metabolic networks are very large and complex systems. For this reason, it is important to build a framework able to unite biology and network theory.  Many successful studies have represented metabolic networks as graphs with metabolites as nodes, reactions as nodes, or both. Taking a step further, with the employment of hypergraphs, we are able to capture what all of these previous graph representations were missing, the higher-order interactions of reactions. In this paper, we show how metabolic networks are naturally mapped into hypergraphs. In particular, the stoichiometry matrix can be viewed as a weighted incidence matrix of a directed hypergraph with edge-dependent vertex weight. No information is lost representing metabolic networks as hypergraphs: the higher-order interactions between metabolites, the directionalities of reactions, and the stoichiometric weights are all included. 

Within this novel framework, we propose two measurements to characterize the hypergraph's robustness and complexity. We apply them to directed hypergraphs with EDVW, but the generalization to undirected and unweighted hypergraphs is straightforward. This approach allows analysis at the local scale, with the communicability and the access and hide information, and at the global scale, with the natural connectivity as a measure of robustness and the average search information as a measure of complexity. We comment on the complications introduced by directionality and how they can be reflected in the measures.
To illustrate the practical application of our framework and metrics, we present an example using the e\_coli\_core model. This small-scale metabolism demonstrates how our metrics operate locally and offers valuable insights into the behavior of metabolic hypergraphs.
At the global scale, we compare 30 different BiGG models in robustness and complexity, leading to some interesting results. We show that the metabolism of organisms that have evolved resistance to antibiotics is associated with hypergraphs that display high robustness. Furthermore, we observe that eukaryotic and prokaryotic organisms have different complexity values. 

A possibility for future works could be modifying the definition of the average search information and the probability of taking a step in the hypergraph. Here, we consider a walk biased by the stoichiometric weights, but more options could be explored. One possibility is to define the probabilities based on the communicability measure or on the rates obtained by flux balance analysis~\cite{FBA_2010,beguerisse-diaz_flux-dependent_2018}. 
Also, we didn't consider the information regarding genes that are contained in the BiGG models. Genomics plays a crucial, especially in resistance to antibiotics, and for this reason, it could be interesting to integrate it into this framework. 
Another possibility is to apply our measures to other contexts, like social or technological hypergraphs.

We believe that this framework represents a promising approach to bridge network theory and biology. We hope that it may serve as a starting point, potentially reaching experts in the field who could further refine and utilize these findings to get more biological insights.

\noindent \\ \textbf{Data availability} \\
All data are publicly available on the BiGG models~\cite{BiGG-models} web page in different formats. In this analysis, the \emph{.json} format is used.

\noindent \\ \textbf{Code availability} \\
Custom code that supports the findings of this study is available from the corresponding author upon request.

\vspace{-0.4cm}
\begin{acknowledgements}
    P.T., G.F.A, and Y.M. acknowledge the financial support of Soremartec S.A. and Soremartec Italia, Ferrero Group. Y.M. acknowledges partial support from the Government of Aragon and FEDER funds, Spain through grant E36-20R (FENOL), and by the EU program Horizon 2020/H2020-SCI-FA-DTS-2020-1 (KATY project, contract number 101017453). We acknowledge the use of the computational resources of COSNET Lab at Institute BIFI, funded by Banco Santander (grant Santander‐UZ 2020/0274) and by the Government of Arag\'on (grant UZ-164255). The funders had no role in study design, data collection, and analysis, decision to publish, or preparation of the manuscript.
\end{acknowledgements}

\vspace{-0.5cm}
\appendix
\section{BiGG models}
In Table~\ref{tab:BiGG_table}, we provide the number of nodes, reactions, and hyperedges for each analyzed hypergraph. We also report the associated BiGG model ID to facilitate its identification, reproduction, and further studies. For more information, see the BiGG models webpage~\cite{BiGG-models}.

\begin{table*}[b!]
    \centering
    \begin{tabular}{|c|c|c|c|c|} \hline
         Organism&  BiGG models&  Metabolites&  Reactions& Hyperedges \\ \hline  \hline 
         Saccharomyces cerevisiae S288C&  iND750&  1059&  1266& 1702\\ \hline 
         Pseudomonas putida KT2440&  iJN746&  907&  1054& 1415\\ \hline 
         Plasmodium cynomolgi strain B&  iAM\_Pc455&  907&  1074& 1563\\ \hline 
         e\_coli\_core&  e\_coli\_core&  72&  95& 141\\ \hline 
         Staphylococcus aureus subsp. aureus USA300\_TCH1516&  iYS854&  1335&  1453& 1872\\ \hline 
         Mycobacterium tuberculosis H37Rv-1&  iNJ661&  825&  1022& 1293\\ \hline 
         Mycobacterium tuberculosis H37Rv-2&  iEK1008&  998&  1224& 1500\\ \hline 
         Clostridium ljungdahlii DSM 13528&  iHN637&  698&  773& 988\\ \hline 
         Yersinia pestis CO92&  iPC815&  1552&  1960& 2507\\ \hline 
 Shigella dysenteriae Sd197& iSDY\_1059& 1888& 2529&3172\\ \hline 
 Escherichia coli str. K-12 substr. MG1655& iJR904& 761& 1075&1329\\ \hline 
 Lactococcus lactis subsp. cremoris MG1363& iNF517& 650& 730&979\\ \hline 
 Helicobacter pylori 26695& iIT341& 485& 554&737\\ \hline 
 Homo sapiens& iAB\_RBC\_283& 342& 469&645\\ \hline 
 Homo sapiens2& iAT\_PLT\_636& 738& 1008&1455\\ \hline 
 Plasmodium falciparum 3D7& iAM\_Pf480& 909& 1083&1576\\ \hline 
 Escherichia coli BL21(DE3)& iEC1356\_Bl21DE3& 1918& 2730&3376\\ \hline 
 Synechococcus elongatus PCC 7942& iJB785& 768& 843&1064\\ \hline 
 Plasmodium berghei& iAM\_Pb448& 903& 1067&1554\\ \hline 
 Trypanosoma cruzi Dm28c& iIS312& 606& 519&806\\ \hline 
 Staphylococcus aureus subsp aureus N315& iSB619& 655& 729&945\\ \hline 
 Thermotoga maritima MSB8& iLJ478& 570& 652&852\\ \hline 
 Methanosarcina barkeri str. Fusaro& iAF692& 628& 690&900\\ \hline 
 Clostridioides difficile 630& iCN900& 885& 1222&1455\\ \hline 
 Plasmodium vivax Sal-1& iAM\_Pv461& 909& 1078&1570\\ \hline 
 Bacillus subtilis& iYO844& 990& 1250&1589\\ \hline 
 Synechocystis sp. PCC 6803& iJN678& 795& 862&1086\\ \hline 
 Geobacter metallireducens GS-15& iAF987& 1109& 1281&1642\\ \hline 
 Acinetobacter baumannii AYE& iCN718& 888& 1013&1436\\ \hline 
 Salmonella enterica& STM\_v1\_0& 1802& 2528&3133\\ \hline
    \end{tabular}
    \caption{BiGG models and their number of metabolites and reactions. The directed hypergraph constructed from the model has a number of nodes equal to the number of metabolites and the number of hyperedges bigger than the number of reactions because of the presence of reversible reactions.}
    \label{tab:BiGG_table}
\end{table*}

\end{document}